\documentstyle[aps,preprint]{revtex}
\begin{document}
\draft
\title{Whirling Waves and the Aharonov-Bohm Effect for Relativistic Spinning
Particles\cite{byline}}

\author{H. O. Girotti}

\address{Instituto de F\'{\i}sica, 
Universidade Federal do Rio Grande do Sul \\ Caixa Postal 15051, 91501-970  -
Porto Alegre, RS, Brazil.}

\author{F. Fonseca Romero}

\address{Instituto de F\'{\i}sica, Universidade de S\~ao Paulo, \\ Caixa
Postal 20516, 01452--990, S\~ao Paulo, SP, Brazil.}

\date{May 1996}

\maketitle

\begin{abstract}
The formulation of Berry for the Aharonov-Bohm effect is generalized to the
relativistic regime. Then, the problem of finding the self-adjoint extensions
of the (2+1)-dimensional Dirac Hamiltonian, in an Aharonov-Bohm background
potential, 
is solved in a novel way. The same treatment also solves the problem of
finding the self-adjoint extensions of the Dirac Hamiltonian in a background
Aharonov-Casher.    
\end{abstract}
\narrowtext

A novel systematics for the computation of the energy eigenfunctions
of a spinless electrically charged nonrelativistic particle in the presence of
an Aharonov-Bohm\cite{AB} external potential, was put forward by 
Berry\cite{Be} sometime ago. We propose here a generalization of this 
systematics to the case where the particle is a relativistic electron. 
As will be shown, the set of eigenfunctions found through this generalization 
serves as a basis for writing any function in the domain of self-adjointness
of the corresponding Dirac Hamiltonian. This corroborates the correctness of 
our proposal, which can also be used to obtain the self-adjoint extensions of 
the Dirac Hamiltonian in an Aharonov-Casher\cite{AC} background field. The 
extension of Berry's ideas to the relativistic regime is the original 
contribution in this work.     

To make this paper self contained, we start by reviewing the main 
features of Berry's formulation\cite{Be}. As known, the Hamiltonian operator 
$H(\vec{r}, \vec{p})$ describing the dynamics of a
spinless two-dimensional nonrelativistic particle (mass $m$ and electric
charge $e$) in the presence of an external static magnetic field $B$ ($B  
= \vec{\nabla} \times \vec A$) is $H(\vec{r}, \vec{p}) = 
H_0\left(\vec{r}, \vec{p} - \frac{e}{c}\vec{A(\vec r)}\right)$,
where $\vec r \equiv (x_1,x_2)$ are the particle coordinates,
$\vec p \equiv (p_1,p_2)$ are the corresponding canonically conjugate momenta 
and $H_0(\vec r, \vec p)$ is the free particle Hamiltonian.
The energy eigenfunctions of the operator $H$
can formally be constructed in terms of those of $H_0$
as follows\cite{Di1}

\begin{equation}
\label{2}
\psi(\vec r)\,=\,\psi_0(\vec r)\,\exp \left( \frac{ie}{\hbar c}
\int_{{\vec{r}}_{0}}
^{\vec{r}} \vec{A}(\vec{r} \prime) \cdot d\vec{r}\prime \right)\,,
\end{equation}

\noindent
where $\vec{r}_0$ is an arbitrary fixed point. The trouble with this
construction is that it leads to multivalued wave functions. Furthermore,
Eq.(\ref{2}) also sais that the magnetic field has no effect on the probability
density $|\psi|^{2}$, which is incorrect. In particular, the vector potential 
of the Aharonov-Bohm background\cite{AB} is 
$\vec{A}(\vec{r})=\Phi \hat{\phi}/2\pi r $ and, correspondingly, 
$B = \Phi \delta(\vec r)$ (we designate by $r$ and $\phi$ the polar 
coordinates and by $\Phi$ the magnetic flux). The computation of the 
magnetic phase factor in the right hand side of (\ref{2}) is, in this case, 
straightforward and one finds that 
$\psi(\vec r)\,=\,e^{ikr\cos (\phi - \theta)\,+\,i \alpha \phi}$,
where $\alpha \equiv e\Phi/2\pi \hbar c$ and $\psi_0(\vec r)$ is a plane wave
whose momentum $\hbar \vec{k}$ makes an angle $\theta$ with the positive
$x_1$-direction ($k \equiv |\vec{k}|$).  For $\alpha$ not an integer, the 
multivaluedness of $\psi(\vec r)$ is explicit.

To remedy this failure, Berry\cite{Be} proposed an alternative procedure for
constructing $\psi(\vec r)$ from $\psi_0(\vec r)$. The new strategy consists of
two steps. First, one recognizes that the partial wave expansion of 
$\psi_0(\vec r)$ ($J_{|l|}(kr)$ is the Bessel function of the first-kind)

\begin{equation}
\label{6}
\psi_0(\vec r)\, = \,e^{ikr\cos (\phi - \theta)}\,
= \, \int_{-\infty}^{+\infty} d\lambda 
\sum_{l = -\infty}^{\infty} (-i)^{|\lambda|} J_{|\lambda|}(kr)
e^{i\lambda (\phi+\pi-\theta)} \, \delta (\lambda - l) \,,
\end{equation}

\noindent
can be rewritten as

\begin{equation}
\label{8}
\psi_0(\vec r)\,=\,\sum_{n=-\infty}^{\infty} T_{n}^{0}(r, \phi)\,=\,
\sum_{n=-\infty}^{\infty} \int_{-\infty}^{\infty} d\lambda (-i)^{|\lambda|}
J_{|\lambda|}(kr) e^{i\lambda (\phi + \pi - \theta + 2n\pi)}.
\end{equation}

\noindent
Secondly, each individual term in (\ref{8}) is magnetized according to the
recipe

\begin{equation}
\label{10}
T_{n}^{0}(r, \phi) \rightarrow T_{n}(r, \phi)\,=\,e^{i\alpha(\phi + \pi -
\theta + 2n\pi)}\, T_{n}^{0}(r, \phi)\,.
\end{equation}

\noindent
One can easily convince oneself that

\begin{equation}
\label{11} 
\psi(\vec r) \equiv\,\sum_{n=-\infty}^{\infty} T_{n}(r, \phi)\,=
\sum_{l = -\infty}^{\infty} (-i)^{|l - \alpha|}
J_{|l - \alpha|}(kr) e^{il(\phi+\pi-\theta)}\,,
\end{equation}

\noindent
which agrees with the expression obtained by Aharonov and Bohm\cite{AB}. 
Observe that the sum in (\ref{8}) is single-valued although the individual 
terms $T_{n}^{0}$ are not, 
$T_{n}^{0}(r, \phi + 2\pi) =  T_{n+1}^{0}(r, \phi)$. The term
$T_{n}^{0}$ has been referred by Berry\cite{Be} as the $nth$ whirling 
wave.

We turn now into generalizing the systematics of Berry\cite{Be} to the 
relativistic regime, which is the main purpose of the present work. The Dirac 
Hamiltonian 
($H^D$) describing the quantum dynamics of a relativistic electron (rest mass 
$m$ and electric charge $e$) under the action of an Aharonov-Bohm potential 
can be written as

\begin{equation}
\label{13}
H^D(\vec r, \vec p)\,=\,H^D_0\left(\vec r, \vec p - \frac{e}{c} \vec A \right)
\,,
\end{equation}

\noindent
where

\begin{equation}
\label{14}
H^D_0\,=\,c\,\epsilon_{ij}\sigma_{i} p_{j}\,+\,smc^2 \sigma_{3}\,,
\end{equation}

\noindent
is the free Dirac Hamiltonian, $\epsilon_{ij}$
denotes the antisymmetric Levi Civita tensor, $\sigma_1$, $\sigma_2$ and 
$\sigma_3$ are the Pauli spin matrices, and $s = \pm 1$.   

The positive energy eigenfunction of $H^D_0$ for $s=+1$ is readily found to 
be

\begin{eqnarray}
\label{15}
 \psi^{0}_{k}(\vec r)\,=\,{\left(\frac{mc^2+E}{2E}\right)}^{\frac{1}{2}}
\left[
\begin{array}{c}
1 \\
\frac{-ic\hbar k e^{i\theta}}{mc^2 + E}
\end{array}
\right] e^{i\vec k \cdot \vec r}\,,
\end{eqnarray}

\noindent
where $E=+\sqrt{m^2 c^4+c^2 p^2}$ is the corresponding energy eigenvalue,
$p\equiv |\vec p|$ and $\vec p = \hbar \vec k$ is the linear momentum of 
the free electron. It will prove
convenient to write its partial wave expansion in the form

\begin{eqnarray}
\label{16}
 \psi^{0}_{k}(\vec r)\,=\,\frac{1}{\sqrt{2E}}\sum_{l=-\infty}^{\infty}
(-i)^{|l|}\, e^{il(\phi-\theta)}\,(-1)^{l}\,
\left[
\begin{array}{l}
 \sqrt{E+mc^2}\,\,\,J_{|l|}(kr) \\
 \sqrt{E-mc^2}\,e^{i\phi} \, \epsilon_{+}(l)\,\,\,J_{|l|+\epsilon_{+}(l)}(kr)
\end{array}
\right]\,,
\end{eqnarray}

\noindent
where $\epsilon_{+}(l) = 1$ if $l \ge 0$ and $\epsilon_{+}(l) = -1$ if 
$l < 0$. By following steps similar to those described in connection with the
nonrelativistic particle, one obtains the whirling wave decomposition of the
upper and lower components of $\psi^{0}_{k}(\vec r)$, namely, 

\begin{eqnarray}
\label{18}
\psi^{0}_{k}(\vec r)& = &\frac{1}{\sqrt{2E}}\sum_{n=-\infty}^{\infty}
\int_{-\infty}^{\infty} d\lambda
(-i)^{|\lambda|}\, e^{i\lambda(\phi-\theta+\pi+2n\pi)}\nonumber\\
& \times & \left[
\begin{array}{l}
\sqrt{E+mc^2}\,\,J_{|\lambda|}(kr) \\
\sqrt{E-mc^2}\,e^{i\phi} \, \epsilon_{+}(\lambda)\,\,
J_{|\lambda|+\epsilon_{+}(\lambda)}(kr)
\end{array}
\right]\,.
\end{eqnarray}

\noindent
Again, one can think of the $n$-th term in each summation of 
(\ref{18}) as 
of a wave arriving at $\phi$ after making $n$-th anticlockwise circuits around 
the origin. 

We next conjecture that the 
effect of adding an Aharonov-Bohm potential is correctly taken into 
account by magnetizing each whirling wave entering in the decomposition 
(\ref{18}). In other words, the 
energy eigenfunction $\psi_{k}(\vec r)$, describing a relativistic 
spinning particle in the presence of an Aharonov-Bohm background potential,
can be obtained from $\psi^{0}_{k}(\vec r)$ through the replacement

\begin{eqnarray}
\label{19}
\psi^{0}_{k}(\vec r) \rightarrow \psi_{k}(\vec r)\,& = &
\,\frac{1}{\sqrt{2E}}\sum_{n=-\infty}^{\infty}
\int_{-\infty}^{\infty} d\lambda
(-i)^{|\lambda|}\, e^{i(\lambda+\alpha)(\phi-\theta+\pi+2n\pi)}\nonumber\\
& \times & \left[
\begin{array}{l}
\sqrt{E+mc^2}\,\,J_{|\lambda|}(kr) \\
\sqrt{E-mc^2}\,e^{i\phi} \, \epsilon_{+}(\lambda)\,\,
J_{|\lambda|+\epsilon_{+}(\lambda)}(kr)
\end{array}
\right]\,.
\end{eqnarray} 

\noindent
This is what we meant in the opening paragraphs of this paper by generalizing
the formulation of Berry\cite{Be} to the relativistic regime. The conjecture 
(\ref{19}) is the main contribution of the present work and its 
meaningfulness will be substantiated by comparing our results with 
those obtained by other authors\cite{Ge,Ha1} for the same
problem\footnote{We are interchanging integrations and 
infinite sums without control of convergence. These formal 
manipulations can not be taken as a substitute for the 
rigorous derivations in Refs.\cite{Ge,Ha1}, without which one could not trust
the results presented in this paper, but as an interpretation of them.}.

By undoing the steps which carried us from Eq.(\ref{15}) to 
Eq.(\ref{18}), one arrives to the following algebraic form for 
$\psi_{k}(\vec r)$ 

\begin{eqnarray}
\label{20}
\psi_{k}(\vec r)\,& = &\,\frac{1}{\sqrt{2E}}\sum_{l=-\infty}^{\infty}
(-i)^{|l-\alpha|}\, e^{il(\phi-\theta)}\, (-1)^{l} \nonumber \\
& \times & \left[
\begin{array}{l}
\sqrt{E+mc^2}\,\,\,J_{|l-\alpha|}(kr) \\
\sqrt{E-mc^2}\,e^{i\phi} \, \epsilon_{+}(l-\alpha)\,\,\,
J_{|l-\alpha|+\epsilon_{+}(l-\alpha)}(kr)
\end{array}
\right]\,.
\end{eqnarray}

\noindent
Two facts can now be established. First, $\psi_{k}(\vec r)$ is a 
single-valued solution for the eigenvalue problem 
$H^D_{+} \psi_{k} = E \psi_{k}$, where $H^D_{+}$ follows from (\ref{13}) 
and (\ref{14}) after setting $s=+1$. Second, any function ($f(\vec{r})$) in
the domain of self-adjointness of $H^D_{+}$ can be written as
$f(\vec{r})=\int d^2k c(\vec{k})\psi_{k}(\vec r)$. We notice that the set 
$\{\psi_{k}(\vec{r})\}$ contains 
functions which are singular at the origin. Indeed, while all partial waves in
the upper component of $\psi^{D}_{+}(\vec r)$ are everywhere regular, the 
partial wave $l = [\alpha]$, in the lower component, becomes singular at the 
origin ($r=0$), as seen from (\ref{20}). Here, $\alpha = 
[\alpha] + \{ \alpha\}$, where $[\alpha]$ is the largest integer $\le \alpha$
and $0 \le \{ \alpha\} < 1$.  Since $l = [\alpha] $ implies that 
$-1 < |l-\alpha|+\epsilon_{+}(l-\alpha) = \{ \alpha \} - 1 < 0$, the singular 
wave is normalizable to a delta function with respect to the measure $r dr$. 
  
The self-adjoint extension we just found is certainly not unique. In fact, 
we can replace (\ref{16}) by the equivalent expansion   

\begin{eqnarray}
\label{21}
\psi^{0}_{k}(\vec r) = \frac{1}{\sqrt{2E}}
\sum_{l=-\infty}^{\infty} (-i)^{(|l+1|+1)} (-1)^{l+1}
e^{il(\phi-\theta)} 
\left[
\begin{array}{l}
\sqrt{E+mc^2} \epsilon_{-}(l+1)
J_{|l+1|-\epsilon_{-}(l+1)}(kr) \\
\sqrt{E-mc^2}\, e^{i\phi}\,J_{|l+1|}(kr)
\end{array}
\right]\,,
\end{eqnarray}

\noindent
where $\epsilon_{-}(l) = 1$ if $l > 0$ and $\epsilon_{-}(l) = -1$ if 
$l \leq 0$. Hence, after magnetizing each whirling wave in (\ref{21}) one
finds that

\begin{eqnarray}
\label{23}
{\psi}^{\prime}_{k}(\vec r)\,& = &\,\frac{1}{\sqrt{2E}}
\sum_{l=-\infty}^{\infty} (-i)^{(|l-\alpha+1|+1)}\,(-1)^{l+1}
e^{il(\phi-\theta)}\nonumber\\ 
& \times &\left[
\begin{array}{l}
\sqrt{E+mc^2}\, \epsilon_{-}(l-\alpha+1)\,\,\,
J_{|l-\alpha+1|-\epsilon_{-}(l-\alpha+1)}(kr) \\
\sqrt{E-mc^2}\,e^{i\phi}\,\,\,J_{|l-\alpha+1|}(kr)
\end{array}
\right]\,,
\end{eqnarray}

\noindent
which is also a single-valued eigenfunction of the operator $H^D_{+}$
corresponding to the eigenvalue $E$. Therefore, another
self-adjoint extension of $H^D_{+}$ has emerged. This time, all the
partial waves in the lower component of ${\psi}^{\prime}_{k}(\vec r)$ are
regular functions of $r$, while the partial wave $l = [\alpha]$, in the upper
component, develops an integrable singularity at $r = 0$.   

The situation for $s = -1$ can be similarly treated.

The last part of this note is dedicated to verify the consistency of our 
proposal for generalizing Berry's formulation\cite{Be} to the relativistic
regime. We shall, then, be comparing the results obtained by us with 
those that already appeared in the literature\cite{Ge,Ha1}.  

In Ref.\cite{Ge} the massive Dirac Equation in an Aharonov-Bohm background
potential was 
solved under appropriate boundary conditions. These conditions were chosen so 
as to secure that the domain of the Hamiltonian was that of its adjoint. The
outcome was a one-parameter family of self-adjoint extensions, whose existence
was confirmed by the method of deficiency indices\cite{VN}. Now, the 
Hamiltonian in Ref.\cite{Ge} is unitarily equivalent to $H^D_{+}$, since 
the unitary matrix

\begin{eqnarray}
\label{26}
U_{+}\,=\,\left[
\begin{array}{cc}
e^{i\pi / 4} & 0 \\
0 & - e^{-i\pi /4}
\end{array}
\right]\,
\end{eqnarray}

\noindent
maps the gamma matrices used by us, for $s = +1$, onto those in 
Ref\cite{Ge}. One can indeed verify that the eigenspaces 
$\{U_{+}{\psi}_{k}(\vec r)\}$ and 
$\{U_{+}{\psi}^{\prime}_{k}(\vec r)\}$ are, respectively, those 
labeled by $\mu = 0$ and $\mu = \pi/4$ in Ref.\cite{Ge}.
\footnote{Recall that the definition of magnetic flux used in Refs.\cite{Ge} 
and \cite{Ha1} differs from ours by a sign.}. 
The point to be stressed here is that, whereas in Ref.\cite{Ge} a judicious 
choice of the boundary conditions is essential for finding the self-adjoint 
extensions of $H^D_{+}$, the method of whirling waves led directly to the 
correct result. We then conclude that the method of whirling 
waves already incorporates the correct boundary conditions, this being true for
the relativistic as well as for the nonrelativistic situations.   

On the other hand,
in Ref.\cite{Ha1} it is first recognized that the operator $({H^D})^2-m^2 c^4$
involves a delta function at the origin. After regularizing this delta
interaction, the solving of the eigenvalue problem yields self-adjoint
extensions of $H^D$ for $s = \pm 1$.
Our gamma matrices representations are unitarily equivalent to those in 
Ref.\cite{Ha1}. For $s=+1$ the mapping is implemented
by ${\bar{U}}_{+} \equiv {U_{+}}^{\dagger}$. One can check that
$\{ {\bar{U}}_{+} \psi^{\prime D}_{+} \}$ coincides with the corresponding 
domain in Ref.\cite{Ha1}.     

To summarize, we have presented in this work a generalization of the
formulation of Berry\cite{Be} which accounts correctly for the quantum dynamics
of a relativistic electron in the presence of an Aharonov-Bohm external
potential. Since the problems are mathematically identical, one may also use
the technique of whirling waves to find the self-adjoint extensions of the
Dirac Hamiltonian in a background Aharonov-Casher\cite{AC,Ha2,Ha1}.


\begin{references}

\bibitem[*]{byline} Supported in part by Conselho Nacional de Desenvolvimento
Cient\'{\i}fico e Tecnol\'ogico, Funda\c{c}\~ao de
Amparo \`a Pesquisa do Estado do Rio Grande do Sul e Funda\c{c}\~ao de
Amparo \`a Pesquisa do Estado de S\~ao Paulo, Brazil. 

\bibitem {AB} Y. Aharonov and D. Bohm, Phys. Rev {\bf115}, 485 (1959).

\bibitem {Be} M. V. Berry, Eur. J. Phys {\bf1}, 240 (1980).

\bibitem {AC} Y. Aharonov and A. Casher, Phys. Rev. Lett. {\bf 53}, 319 
(1984).

\bibitem {Di1} P. A. M. Dirac, Proc. R. Soc. {\bf A133}, 60 (1931).

\bibitem {Ge} Ph. de Sousa Gerbert, Phys. Rev. D{\bf 40}, 1346 (1989).

\bibitem {Ha1} C. R. Hagen, Phys. Rev. Lett. {\bf 64}, 503 (1990);
Int. J. Mod. Phys. A{\bf 6}, 3119 (1991).

\bibitem {VN} M. Reed and B. Simon, {\it Fourier Analysis and Self-Adjointness}
(Academic Press, New York, 1975).

\bibitem {Ha2} C. R. Hagen, Phys. Rev. Lett. {\bf 64}, 2347 (1990).

\end{references}
\end{document}